# ICT in Universities of the Western Himalayan Region of India II: A Comparative SWOT Analysis

Dhirendra Sharma[1] and Vikram Singh[2]

[1]University Institute of Information Technology, Himachal Pradesh University,
Shimla, Himachal Pradesh 171 005, India

[2]Department of Computer Science and Engg, Ch. Devi Lal University,
Sirsa, Haryana 125 055, India

**Abstract**
This study presents a comparative SWOT analysis to comprehend the pattern of development of ICT within six universities of western Himalayan region of India. With the objective of achieving quality and excellence in higher education system in the region, this study provides a basis to decision makers to exploit opportunities and minimize the external threats. The SWOT analysis of different universities, placed under three categories, has been undertaken within the four-tier framework used earlier by the authors. Guided by the initiatives of National Mission on Education through ICT (NMEICT) for SWOT analysis, findings of this paper reveal, relative consistency of these three categories of universities, with the earlier study. A few suggestions, as opportunities, with an emphasis on problem solving orientation in higher education, have been made to strengthen the leadership of universities in the field of ICT.

## 1. Introduction

SWOT (Strength, weakness, opportunities and threats) analysis has proved to be a general tool at the preliminary stages of policy making and strategic planning of an organization and at a later stage as well, while analyzing the performance and planning for further development and progress of the organization. For the latter, the SWOT analysis serves double purpose of getting the answers to some very relevant questions from the performance and planning for future development, success and failure/ difficulties faced by the universities/organization. In view of that experience, one looks for opportunities provided by such an analysis. A saying is that opportunities knock the door at least once which is to be promptly identified and utilized.

In this connection, Narayana Murthy [11], non-executive chairman of Infosys Technologies, Bangalore, India, while addressing a convocation in Jammu, admonished that problem- solving oriented education system should be the priority which would produce skilled professionals as good teachers and should be adopted by the universities as their sacred responsibility. Only then the challenges faced by the country due to widening of the gap between the haves and have-nots, arising from the technological development, may be bridged.

### 1.1 SWOT as an Analytical Technique

The origin of the SWOT, as an analytical technique, lies in connection with the growth of strategic planning which dates back to the decade of 1960s. The concept was developed later, to address possible shortcomings in the outcome of strategic planning [2,9,10].

SWOT has established itself as a framework for analyzing strengths, weaknesses, opportunities and threats. Strengths and weaknesses are mainly based on internal audit, as a result of introspection of a universities/organization. The opportunities are related with the internal as well as external environmental factors. Threats are concerned mainly with the external environment factors. The external factors imply economy, competition, sources of funding, demographics and culture. These are needed to be taken care in strategic planning and activities. Opportunities represent factors that can be beneficially exploited. Threats need to be considered because of the potential of damaging the organization/ institute. SWOT analysis normally reflects a viewpoint which can be used by others. It has to be positive so that the analysis is exploited for the benefit of the organisation. Different variants of SWOT [6,16] provided a structure based planning [3,14] and implementation. This technique is used to develop a project or find a solution to a problem that takes care of several different internal and external factors and maximizes the potential of strengths and opportunities while minimizing the effect of weakness and threats.
In a recent paper [5], presented a comparative SWOT analysis of four universities in the pacific Asian region in the context of information system. Supported by their historical perspective, they elaborated on policies, their implementation and strategic management in those universities. The historical perspective gives a bird's eye view of evolution of the university in ICT field, which can be quite revealing. A retrospective analysis of this perspective may be useful in documenting key changes over time. With this, the SWOT analysis may provide directions to assist in making decisions and strategies about the relative merits of different activities in the ICT universities.

In order to undertake a SWOT analysis with rigor, an essential pre- requisite is that the primary data collected should be through persons who have a deep understanding of the organization, including its historical perspective. This would enable one to identify its strengths, weaknesses and opportunities as well as a sound understanding of internal and external environment, which may effect positively as opportunities and negatively as harmful effects and threat [8].

### 1.2. National Mission on Education through ICT

One of the most crucial challenges facing Indian higher education is its quality for which Government of India has an ambitious goal for eleventh plan. Recently, Government of India through Ministry of Human Resource Development (MHRD) has developed a holistic approach on National Mission on Education through ICT (NME ICT)[15]. NMEICT has brought out a document, which has already been triggered during the period of tenth plan (phase I). As per its strategy, its future vision, planning and developmental activities will form phase II and phase III during the eleventh five year plan period.

It has an ambitious vision of providing one stop solution for the learning community. The working document of the mission is concerned with the education from school/ college (regular & engineering) level to university level. It has three guiding principles.

**Human resource development:** Talent in the higher education should be identified, trained and utilized in the service of the country.
**E- content/ resource development.** Quality e- content should be developed and delivered through the network connectivity of NME ICT.
**Building connectivity and knowledge network**: In order to provide maximum benefit to the learners, the maximum possible inter- connectivity should remain available among and within institutions of higher learning in the country with a view to achieve critical mass of skilled human resource/ researchers in any given field.

These guiding principles are expected to lead to various important steps in planning and implementation as follows:

- ICT Technology should reach to each learner
- Generation of quality e-content, questions bank as modules-based learning.
- Development of interface modules for physically challenged learners.
- Facility of Geographical Information System (GIS) for planning upto the village level.
- Improvement in course curriculum and teachers training programs.
- Efficient and effective knowledge transfer to learner with proper interaction
- Voice over Internet Protocol (VOIP) supported communication between learner and teacher
- Enterprise Resource Planning (ERP) and e-governance for education, coordination & synergy for implementation of the policies, setting up virtual laboratories and support for creation of virtual technical universities.
- Performance optimization of e-resources
- Certification of attainments of any kind at any level.

All these factors are supposed to contribute towards the SWOT analysis of any higher educational institution.
In the present paper SWOT analysis will be carried out through a 2x2 matrix worksheet as given in Table I, of ICT in six universities of western Himalayan region of India.

Table I

| Strengths | Weakness |
|---|---|
| Opportunities | Threats |

Such a work sheet for SWOT analysis is particularly suited in providing a structure, objectivity, transparency, if it is carved out by a small visionary group. Strengths and weaknesses may be visualized from the success and failure of an organization at the level of implementing policies and its performance thereafter. Further, one has to identify promptly the most attractive opportunities arising from internal factors to convert the weaknesses/ failures and then the external environmental factors, to the advantage of the organization. In general, internal weaknesses must be tackled first, in the form of opportunities, before looking at the external environmental factors.

It is always advisable to analyses these four ingredients of SWOT in a systematic manner. One of the ways, the authors feel, is to follow a four- tier framework containing vision and planning, infrastructure, activities, performance & impact. Higher is the aim & the mission more intense is the vision. Accordingly the infrastructure is acquired, which is quite easy, given the financial resources. However, it is really a passive ingredient without the skilled human resource. The real strength of the organization lies in programming and organizing the activities more and more efficiently using the skilled human resource which is only possible through the values of self- discipline and dedication that inculcates a work culture in them. Without these values one can not dream of quality in higher education, howsoever good the planning or the infrastructure may be. These value- based qualities in skilled human resource provide strength to the organization in the form of better performance and their dilution makes the organization weaker and weaker. The better performance through its product, in turn, leads to have an impact at the national and international level. For the university system the four- tier framework is given below.

## 1.3. Four- Tier Framework

In a recent paper Sharma and Singh (2009) presented, a detailed analysis of initiative & planning, status and performance, obtained from the primary data based on the questionnaire, in the field of ICT in the universities of western Himalayan region of India. The analysis was carried out within a four– tier framework, Fig.1, containing vision & planning, infrastructure, activities and performance. This study will form the basis for the SWOT analysis discussed here.

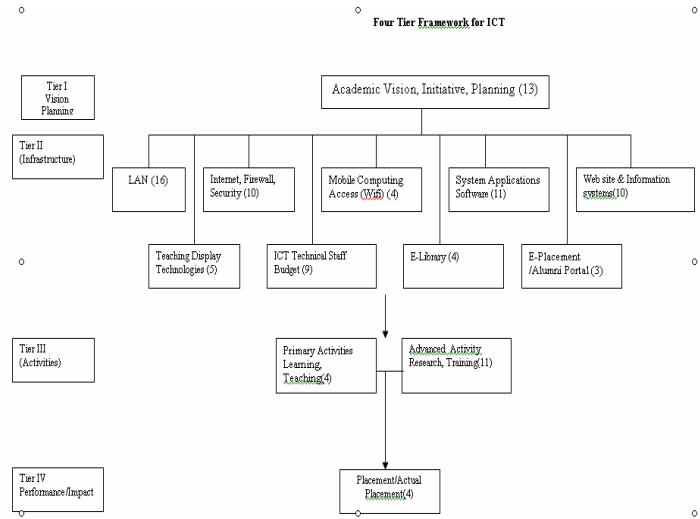

Fig.1. Four Tier Framework for ICT. The number in bracket denotes the number of questions related to the item/ group in the questionnaire.

The focus of that analysis was on the performance assessment based on the vision & planning, dynamic curriculum, good infrastructure with prompt technical support, technically skilled professionals & better interaction with the students and integration of ICT with all its activities particularly the problem solving ability at the academic level. It was further emphasized that universities are meant to create knowledge to operate at the universal level where quality remains the defining element. According to that study six universities of this region are placed in three categories as follows (Table II).

Table II: Details of six universities in three categories, A, B and C

| Category | University | Abbreviation | State | Founded Year | Ownership | Website |
|---|---|---|---|---|---|---|
| A (Technical) | J.P.University of Information Technology, Solan, H.P. | J.P.U, Solan | Himachal Pradesh | 2002 | Private | http://www.juit.ac.in |
| | National Institute of Technology (Deemed University) Hamirpur. | NIT, Hamirpur | Himachal Pradesh | 1986 | Central Govt. of India. | http://www.nitham.ac.in |
| B (Agriculture) | C.S.K.Agriculture University, Palampur | C.S.K.A.U., Palampur | Himachal Pradesh | 1978 | State Government | http://hillagric.ernet.in |
| | Y.S.Parmar Horticulture University Nauni, Solan | Y.S.P.H.U. Nauni | Himachal Pradesh | 1985 | State Government | http://www.yspuniversity.ac.in/ |
| C (Regular) | H.P.University, Shimla | H.P.U,Shimla | Himachal Pradesh | 1970 | State Government | http://www.hpuniv.gov.in |
| | Jammu University, Jammu | J.U.Jammu | Jammu and Kashmir | 1969 | State Government | http://www.jammuuniversity.in/ |

The paper is organized in five sections as follows. Having presented the introduction in section 1, the SWOT analysis of technical universities will be presented in section 2. Section 3, deals with the SWOT analysis of agriculture and horticulture universities. Section III will be devoted to regular multi-faculty universities and their SWOT analysis. The conclusion will be given in the final Section.

## 2. Technical Universities:

There are two technical universities out of the six selected from the western Himalayan region. One is J.P.University of Information Technology, Solan and another is National Institute of Technology, Hamirpur (Deemed University).

2.1. Historical perspective:
Historical perspective of J.P.University and NIT Hamirpur are presented in Table III A and IIIB respectively.

Table III A. Historical Perspective of J.P. University

| Faculty | Department | IT Course | Year |
|---|---|---|---|
| Faculty of Engineering | Computer Science & Engg, Information Technology, Electronics and Communication, Civil, Bio-Informatics | B.Tech | 2002 |
| | | M.Tech, Ph.D | 2006 |

Table III B. Historical Perspective of NIT Hamirpur

| Faculty | Department | IT Courses | Year |
|---|---|---|---|
| | Computer Centre | | 1986 |
| Faculty of Engineering | Electrical and Civil. | B.Tech, | 1986 |
| | Electronics & Communication | B.Tech, | 1988 |
| | Computer Science | B.Tech, | 1989 |
| | Mechanical | B.Tech, | 1993 |
| | Architecture | B.Arch. | 2000 |
| | Electrical, Civil, Electronics & Communication, Computer Science, Mechanical | M.Tech and Ph.D | 2006 |
| | Management and Social Science. | MBA, Ph.D | 2009 |

Both of these technical universities have single faculty of engineering with different engineering departments including computer science, electronics and communication and have been running B.Tech, M.Tech and Ph.D programs.

J.P.University in addition has a department of Information Technology and Bio-informatics. These programs have been going on in respective departments since 2002, whereas in NIT Hamirpur, computer centre was established in 1986, department of electronic and communication was started in 1988 and computer science department was setup in 1989. The department of management came in existence in the year 2008.

In general, at both the universities the courses in computer science are compulsory to learn the ICT skills, for all the students belonging to different branches of engineering. But emphasis on the ICT programs is given through the curriculum in the departments of computer science, electronics and communication, information technology and bio-informatics. The department of management in NIT Hamirpur has also been conducting courses on information system.

2.2 SWOT Analysis:

SWOT analysis of these two technical universities (J.P.University and NIT Hamirpur) is given in Table IVA and IV B.

Table IV A. SWOT Analysis of J.P. University, Solan

| Internal strengths | Internal weakness |
|---|---|
| *Vision-* Initiative and ICT Planning<br>*ICT Infrastructure*<br>Network internet and security, Information system, ERP, Teaching / technical staff, ICT budget allocation, E-library system/ E- Content, E-placement/alumni portal, ICT based teaching and learning, Redundancy feature in firewall,<br>*Activities-*Greater Industry Interaction.<br>*Performance-* Research, Placement. | *Vision-*<br>ICT Governance, Leadership<br>*ICT Infrastructure-*<br>Mobile computing. Video-Conferencing Location specific, lack of qualified faculty<br>IP Telephony<br>Lack of redundancy feature in campus wide network backbone.<br><br>*Activities-*<br>Weak student teacher interaction |

| External Opportunities | External Threats |
|---|---|
| Effective interaction with Alumni and industries. Collaboration with universities and industries. Entrepreneurship programs Solution to environmental disasters. Development of skilled professionals. | Government policy and norms Growing digital divide. Threat from other private /foreign universities players. |

Table IV B. SWOT Analysis of NIT Hamirpur

| Internal strengths | Internal weaknesses |
|---|---|
| *Vision-* Effective Leadership & ICT Governance. *ICT Infrastructure* LAN Wired and Wireless, Effect website, Internet firewall security, video Conferencing, IP Telephony within campus, Maintenance of networks, Informative website, IS, ERP, Alumni portal/association, ICT technical staff, Maintenance, ICT budget, E library, E- content, Well qualified faculty *Activities* Problem based teaching & learning, Greater industry interaction. *Performance* Research collaboration, Actual placement, Training for faculties. | *ICT Infrastructure-* Lack of redundancy feature in campus wide network backbone and firewall. |
| **Opportunities** | **Threats** |
| More ICT based training programs for professionals. More effective contact with alumni. Develop new ICT tools for teaching and e-learning | Threat from security, government policy, Private foreign universities. |

**Strengths:** As per our general framework, the vision and ICT planning and various initiatives belonging to tier I, are reflected through good ICT infrastructure, networking, internet security, information system/ ERP, e-library system, effective websites and video conferencing facility. Both the universities are having good maintenance network and of computers, close academic industry interaction, e-placement and alumni association/portal. It is interesting to point out that NIT Hamirpur has an edge over J.P.University because of:

a) The availability of best financial resources from Government of India (GOI).
b) NIT Hamirpur draws better faculty and technical staff in sufficient number. They are encouraged to improve their technical skills and qualification at the national level through quality improvement programs organized by AICTE.
c) NIT Hamirpur has better facility of IP telephony, wireless network and student counseling.
d) NIT Hamirpur organizes application and training/ extension programs using ICT facilities, for the faculty of different universities and engineering colleges.
e) Better ICT facilities including the concentration on problem- based learning makes teaching programs better, improving the quality of students.

Both the universities, due to greater industry interaction, have better opportunities of placement for outgoing students at the national and international level. Nevertheless, J.P. University encourages students to join J.P group itself. In overall performance, NIT Hamirpur has an edge over the other.

**Weaknesses:** Both the technical universities lack in campus wide network backbone redundancy features. NIT Hamirpur also lacks in redundancy of firewall. J.P. University is supposed to be weak in student- teacher interaction due to the large number students in the class room as compared to NIT Hamirpur. Whereas J.P.University lacks in mobile computing, IP telephony and Video-conferencing.

**Opportunities:** Opportunities can be divided into two groups; one coming from internal factors and another from external/environmental factors. Most of the internal weaknesses of an institution can become attractive opportunities manifested mainly in the form of performance.

In this respect, redundancy feature of the campus wide network backbone, close contact with its alumni which J.P. University may be able to handle in a better way, are some of the opportunities for both of them. In particular these are the alumni who always prove very helpful to the organizations and for providing placement to outgoing students. These technical universities are also having a greater responsibility towards finding solutions to environmental disasters and development of skilled professionals along with the development of entrepreneurs.

**Threats:** Being financially sound, there is no internal threat as such at the level of ICT infrastructure and activities. At the performance level both the institutions have a threat from the security in social context, which may be external. Further, there may be a threat from private/ foreign universities in future. Growing digital divide, again in social context, may be another threat. For a private university like J.P.University there may be threat from the government policies framed from time to time.

## 3. Agriculture and Horticulture Universities:

In this category there are two universities. One is Agriculture University, Palampur and another is Horticulture University, Nauni, and Solan. Initially, Palampur campus for agriculture and Nauni campus for Horticulture/ Forestry were part of the regular multi-faculty, Himachal Pradesh University, Shimla, which trifurcated in the year 1977 resulting in a full fledged Agriculture University Palampur and Horticulture University, Nauni, Solan

### 3.1 Historical Perspective.

The historical perspective of Agriculture University, Palampur and Horticulture University, Nauni, Solan are displayed in Table VA and Table VB, respectively

Table V A. Historical Perspective Agriculture University, Palampur

| Faculty | Department | IT Courses | Year |
|---|---|---|---|
|  | Computer Centre (Central Facility) | UG and PG Level | 1988 |
| Agriculture, Veterinary & Animal Science and Physical Science | Agriculture, Veterinary & Animal Science and Physical Science | Fortran Programming | 1991 |
|  |  | Introduction to computers and ICT, C Programming | 2000 |

Table VB. Historical Perspective of Horticulture University, Nauni

| Faculty | Department | IT Courses | Year |
|---|---|---|---|
|  | Computer Centre (Central Facility) | UG and PG Level | 1988 |
| Horticulture & Forestry | Horticulture & Forestry | Computer Science and Computer Applications |  |
| Agriculture Business Studies | Agriculture Business Studies | Introduction to Computer, MIS and Computer Applications in MBA. | 1997 |

The computer centres as central facility, was established in both the universities in 1988. Since then they offer compulsory courses on professional skills on computers and its applications including programming, in each branch of studies at the UG and PG level.

Horticulture University, Nauni was the first to have network facility and internet connectivity provided by ICAR in the year 1996. VSAT was commissioned in the year 1999 and OFB was laid in the following year. Horticulture University, Nauni has two faculties i.e. the faculty of Horticulture & Forestry and the faculty of agriculture business studies. In the faculty of agriculture business studies, the course on computer application was started in the year 1997 for MBA students. They also organize program on Geographic Information System (GIS) frequently using ICT facilities. Whereas, agriculture university, Palampur has three faculties namely the faculty of agriculture, the faculty of veterinary & animal science and the faculty of physical science. Both these universities organise extension programs with the help of ICT infrastructure.

### 3.2. SWOT Analysis.

SWOT analysis of these two universities is given in VIA and VI B.

Table VIA. SWOT Analysis of Agriculture University, Palampur

| Internal strengths | Internal weaknesses |
|---|---|
| *Vision*-ICT planning | *Vision*- ICT Motivated leadership |
| *ICT Infrastructure-*   LAN facility   Internet and security   Mobile computing   E library | *ICT Infrastructure-*   Lack of redundancy feature in campus backbone.   ICT support system.   IS and ERP   ICT Budget   Maintenance of computers |
| *Activities:*     ICT based teaching and training,   Extension programs | *Activities*     E placement and alumni portal   ICT technologies in teaching |
| *Performance:*   Information from effective Website.   Research. |  |

68

| Opportunities | Threats |
|---|---|
| Effective University Industry Interaction Close Contact with Alumni. Effective Research Collaboration Bandwidth upgradation | Threat from government policies. Threat from other private players. |

Table VI B. SWOT Analysis of Horticulture University, Nauni.

| Internal strengths | Internal weaknesses |
|---|---|
| *Vision*-ICT Planning **ICT infrastructure** LAN facility E-library system *Activities-* ICT based teaching, learning Training/Extension programs *Performance-* Actual placement. | *Vision-* ICT Motivated Leadership **ICT Infrastructure-** Internet, security, redundancy in OFB IP Telephony, Video Conferencing, Mobile computing, Information system and ERP, ICT Teaching technologies, Budget allocation, ICT support system/Placement Portal, Teaching and Technical Staff *Activities-* Research using ICT |
| **Opportunities** | **Threats** |
| Close Contact with Alumni. More Problem oriented programs. Bandwidth upgradation | Powerful private and foreign Universities. Government policies. Network is not secured. |

**Strength:** Both these universities share similar vision and planning in their respective fields of agriculture, horticulture and forestry. The ICT infrastructures like network, internet connectivity, e-library system are also similar and both of them organize extension programs in order to train the professionals. These two universities have exclusive academic faculties in respective disciplines; as a result they have independent programs and activities.

Horticulture University, Nauni has an additional faculty of agriculture business studies in which they have special feature of organizing activities related to agricultural business management as per industry requirement. They also train professionals in the field of Geographical Information System (GIS), an application oriented program which gives information upto the village level. Due to this innovative program the students from this university have better placements. Further, they also conduct various special courses, such as floriculture, sericulture, mushroom cultivation. These activities are supported by business management skills which are quite advantageous to the students in being good entrepreneurs.

Agriculture University, Palampur has a relative advantage

| Faculty | Department | Courses | Year |
|---|---|---|---|
| Faculty of Physical Science | Computer Centre | CIC, DCA | 1987 |
| | Computer Science | MCA, PGDCA | 1989 |
| | ICDEOL | APGDCA | 2004 |
| | ICDEOL | PGDIT | 2005 |
| | Computer Science | M.Tech.(CS) | 2006 |
| Faculty of Engg & Technology | UIIT | B. Tech. (IT) | 2000 |
| Faculty of Commerce & Mgt | Institute of management of Studies. | MBA (Information Systems) | 2002 |

of their better informative, effective and functional website with internal mail system. They have also provided the internet connectivity with better backbone on the campus.

**Weakness:** These two universities are lacking in information system, ERP, E-placement and alumni portal, redundancy feature in campus backbone and firewall. They also lack in e-library system and e-content management, IP telephony, video-conferencing, mobile computing. Being state universities, both the universities are having financial constraints.

Though Horticulture University, Nauni is better in planning and ICT infrastructure, however, it got leveled due to weakness in activities and performance as compared to that in agriculture university, Palampur, which organises better ICT activities through more informative web site and intranet services.

**Opportunities:** The above weaknesses may be converted into opportunities for development. Both the universities have opportunity in the form of upgrading the bandwidth and other desired infrastructure alongwith the more activities.

**Threats:** The network is not secure, in social context, in either of the universities. There may be a threat from private and foreign players.

## 4. Regular Multi-faculty Universities:

There are two universities in this category. These are Himachal Pradesh University, Shimla and Jammu University, Jammu.

### 4.1 Historical perspective-

Historical perspective of these two universities is given in VIIA and VII B respectively:

Table VIIA. Historical Perspective of H. P. University, Shimla

Table VII B. Historical Perspective of Jammu University, Jammu

| Faculty | Department | Courses | Year |
|---------|------------|---------|------|
| Physical Science | Computer Centre (Central facility) |  | 1987 |
| | Computer Science | Diploma in Computer Science | 1987 |
| Physical Science | Computer Science | MCA | 1995 |
| Physical Science | Computer Science | Ph.D |  |
| Management | Management Studies | PG Diploma in Mgt Studies |  |
| | Management Studies | MBA |  |

Himachal Pradesh University, Shimla started functioning in 1971 and had been running various PG programs in more than a dozen faculties including faculty of agriculture and horticulture/forestry. As a result of trifurcation in the year 1977, full fledged agriculture university at Palampur and Horticulture University at Nauni, Solan, came into existence in the state of Himachal Pradesh. Presently, Himachal Pradesh University is having more than 30 teaching departments, on its campus.

The Computer centre at Himachal Pradesh University, Shimla was established 1987, under faculty of physical science, with diploma course in computer applications. In the year 1989, MCA program was started and DCA was upgraded to PG diploma in computer applications (PGDCA). In the year 2004, VSAT connectivity (512 kbps) was installed with bouquet of more than 4000 ejournals, for academic community of the university. This facility was centrally located on the university campus.

In the year 2007, terra-byte optical fibre backbone connectivity was commissioned. All the teaching faculty members, various teaching laboratories and administrative officers got the internet facility right in their offices. Total number of users on campus became 810. This internet facility provided access to all e-journals through the Inflibnet (INFLIBNET), for teachers, researchers of this university. In the year 2008, connectivity of 512 Kbps was upgraded to 2 Mbps (1:1) leased line. The ICT infrastructure developed on the campus is being used by whole of the university.

B.Tech. program in information technology was started in the year 2000, with the University Institute of Information Technology. MBA (Information System) was triggered in the year 2002 under the faculty of commerce and management. M.Tech. in computer science began in the year 2006, under the faculty of physical sciences.

University of Jammu came into existence in 1969. Computer centre was established in the year 1987. University Optical Fibre backbone was established in 2003. In the year 2005, a comprehensive website of this university became functional. The DCA program was launched in 1987 followed by MCA program in 1995 under the department of computer science. The department of management studies has been conducting MBA (IT) program on the campus. It is pertinent to mention that whole of the academic community is being benefited by these ICT facilities.

### 4.2 SWOT Analysis:

Himachal Pradesh University and Jammu University are state universities of the state of Himachal Pradesh and Jammu & Kashmir respectively in the western Himalayan region. As a result, they have constraints on the financial resources. Both are multi-faculty universities and are working for diverse disciplines. The SWOT analysis of these two universities is given Table VIII A and VIII B.

Table VIII A. SWOT Analysis of H. P. University, Shimla

| Internal strengths | Internal weaknesses |
|--------------------|---------------------|
| *Vision:* ICT Planning  *ICT Infrastructure*  LAN Facility  Internet and firewall Security, | *Vision* – ICT oriented leadership  *ICT Infrastructure-*  Non availability web server/mailserver/e-content delivery system/DNS Services. Redundancy feature in firewall. Videoconferencing IP Telephony Mobile computing(Wireless) ERP and E- |
| Redundancy feature in OFB. *Activities:* Training/Extension programs(ASC) *Performance* using ICT Research | Governance ICT Technologies for teaching Qualified & sufficient Technical Staff Maintenance of ICT infrastructure ICT budget E-library system/automation/ e-Contents. *Activities-* E-placement and alumni association. Problem based learning/teaching approach. Collaboration with other universities. Synergy with multi disciplinary activities Training of faculty/skilled professionals, Problem oriented education. |
| **Opportunities** Collaboration with industries. Entrepreneurship Solution of ICT disaster. Close contact with alumni | **Threats** Migration of students to other universities, Presence of private foreign universities. Threat from government policies. |

Table VIII B: SWOT Analysis of Jammu University, Jammu

| **Internal strengths** | **Internal weaknesses** |
|---|---|
| *Vision*- ICT Planning **ICT Infrastructure** Impact of ICT, research and placement. LAN facility Internet and firewall security Mobile Computing Activities- Extension programs(ASC) Performance Research | *Vision* Motivating Leadership. *ICT Infrastructure*- Redundancy feature in backbone and Firewall, video conference facility IP Telephony, IS/ ERP ICT Technologies in Teaching, ICT teaching, technical staff, E-library system, E content, ICT support System *Activities* – Industry university of interaction Research Performance. Problem Oriented Training/Faculty |
| **Opportunities** Close Contact with Alumni. University Industry Interaction. Collaboration with Foreign universities | **Threats** Wi-fi is not secured. Threat from other private players. Threat from government policies. |

**Strength**: In vision and planning, Himachal Pradesh University had a little advantage over Jammu University in ICT infrastructure, networking and security; both the universities are at par. In addition to this, Jammu University is having its own Web server, mail server, DNS facility, better mobile computing, and more internet bandwidth as compared to Himachal Pradesh University

**Weakness:** Himachal Pradesh University is lacking with the facility of web server, mail server, DNS facilities, and video conferencing. Mobile computing facilities, ERP, e-content, e-governance are not available in both the universities. They lack sufficient ICT technologies for teaching, e-library, e-placement, ICT support system, maintenance of computers along with sufficient well qualified teaching faculty. Nevertheless, Jammu University has certain advantage in respect of video conferencing and better bandwidth connectivity over Himachal Pradesh University.

**Opportunities:** The weaknesses in respect of ICT are to be converted into opportunities in infrastructure and activities. The more crucial is to adopt the technology of problem solving orientation in learning as internal factors and to face the challenges due to external factors like collaboration with other universities, industries and development of ICT applications at the advanced and professional level. Establishment of close contact with alumni will also be helpful for both the universities.

**Threats:** Threats come from government policies, private & foreign universities. ICT security is the major threat in both the universities.

## 5. Conclusions and Suggestions

We have presented a comparative SWOT analysis in respect of ICT, of six universities placed in three categories, supported by their historical perspectives. This has been done within the four- tier framework of ICT[15] and on the basis of primary data/ feedback obtained from different universities. Findings of this paper are along the lines to those of National Accreditation and Assessment Committee (NAAC), an autonomous body of University Grants Commission as far as regular multi faculty universities are concerned. ICT activities have a crucial role to play as per NMEICT directions/ policies to be adopted by the universities in order to achieve quality and excellence in higher education system in the region.

On the basis of this SWOT analysis, answers to some of the glaring questions regarding ICT ingredients may be briefly mentioned as follows:

*ICT Vision and Planning (TIER I):*
- A motivating ICT oriented leadership should be provided to universities.
- There should be academic flexibility with a dynamic curriculum through the feedback obtained from alumni, teachers and industries.
- Interdisciplinary courses based on ICT should be encouraged.

*ICT Infrastructure (Tier II)*
- Adequate internet bandwidth and number of e-journals should be made available.
- The network system should be best secured against anti- national elements.
- Latest technology should be adopted from time to time.
- ERP systems are implemented so as to have paperless communication along with an interface with the academic community.
- There should be ICT based connectivity among various universities and institutions so as to share knowledgeable e- resources.

*Activities (Tier III)*
- ICT has yet to make an impact in classrooms.
- Problem- solving oriented education system is followed in the universities, so as to train skilled professionals, providing responsible leadership through qualified and dedicated teachers.
- Interaction with alumni (with a strong data base) and industries needs to be enhanced.
- All university administrative wings be computerized and integrated with ERP.
- Faculty members are motivated to take the projects/ consultancy.
- More effective counseling is made available to the academic community.

*Performance (Tier IV)*
- Internal Quality Audit Cell (IQAC) should be strengthened for quality enhancement
- University should explore placements through their alumni and corporate sector.
- Collaboration among different universities/ institutions in order to share e- resources and research projects at the national and international level.
- More and more collaboration with industries be encouraged.
- Classroom teachings are strengthened through ICT.

There seems to be a downturn in IT field which may affect all the universities alike and should be taken as a challenge. This can be overcome by improving the quality of teaching through problem solving orientation/ training programs as per the need of the country and the global scenario.

All these suggestions may be seen as opportunities for the development of the system of various higher educational institutions/ universities.

**Acknowledgments:**

The authors thank Dr. S.P. Saraswat, Agro- Economic Centre, Himachal Pradesh University, Shimla for his valuable discussions during the SWOT analysis.

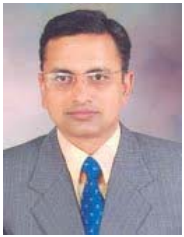

**Dhirendra Sharma** has obtained his MBA (1998) from Maastricht School of Management, Maastricht, Netherlands, M.S (Software System) (1996) from BITS, Pilani, India, M.Sc.Physics (1989) & M.Phil. (Physics) (1990) from Himachal Pradesh University, Shimla, India. His areas of interest are ERP and its implementation in higher educational institutions, computer networking (wired and wireless), Wireless Sensor Networks (WSN), and open source web content management. He played a very important role in Design and Implementation of Campus Wide Optical Fibre Network at Himachal Pradesh University, Shimla. He is having more than 10 years of teaching experience in addition to his 5 years in IT Industry. He is having more than 07 publications in international/national journals/conferences. At present he is pursuing Ph.D. from the Department of Computer Science and Engineering, Ch. Devi Lal University, Sirsa, Haryana, India.

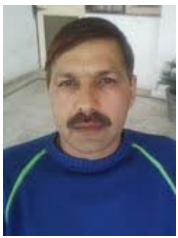

**Dr. Vikram Singh** is Ph.D (2004) in Computer Science from Kurukshetra University, Kurukshetra, India. Presently he is working as Professor and Head in the Department of Computer Science & Engg and Dean Faulty of Engineering, Ch. Devi Lal University, Sirsa – 125055. Haryana, (India), since 2004 onwards. Earlier he was working with Kurukshetra University, Kurukshetra. His areas of interest are computer networks, e-Governance, and system simulation tools. He is having more than 17 years of teaching/ research experience, has more than 30 publications in international and national journals/conference proceedings, alongwith three books on the subject.